\documentclass{article}  
\usepackage{lajolla2006}
\usepackage{graphicx}
\usepackage{amsmath,amstext,amsthm,amssymb,amsfonts}
\frompage{000} \topage{000}                                                       
\def\rightmark{Nonequilibrium dilepton production}
\def\leftmark{C. Greiner et al.}
 
\title{Dileptons from vector mesons with dropping masses\\
in a nonequilibrium quantum field theoretical framework} 
\authors{
{C. Greiner and B. Schenke%
}\\[2.812mm]
{\normalsize
\hspace*{-8pt} Institut f\"ur Theoretische Physik\\ \hspace*{-8pt} Johann Wolfgang Goethe - Universit\"at Frankfurt\\
\hspace*{-8pt} Max--von--Laue--Stra\ss{}e~1, D--60438 Frankfurt am Main, Germany\\[0.2ex] 
}}
 
\abstract{The influence of time dependent medium modifications of low mass vector mesons on dilepton production
is investigated using nonequilibrium quantum field theory.
By working in the two-time representation memory effects are fully taken into account. 
For different scenarios the resulting dilepton yields are compared to quasi-equilibrium calculations and 
remarkable differences are found leading to the conclusion that memory effects can not be
neglected when calculating dilepton yields from heavy ion collisions. Emphasis is put on a dropping mass
scenario since it was recently claimed to be ruled out by the NA60 data. Here the memory effects turn out to be particularly
important.}

 
\begin{document}
 
\maketitle
\setcounter{page}{1}

\section{Introduction}\label{intro}

Relativistic heavy ion reactions allow for studying strongly
interacting matter at high densities and temperatures. 
One of the main objectives is the creation and
identification of new states of matter, most notably the
quark-gluon plasma (QGP). Photons and dileptons do not undergo
strong interactions and thus carry undistorted information
on the early phases of the fireball because the production rates increase rapidly with temperature.
Photon spectra are a suitable observable for the temperature
whereas in the low mass region dileptons couple directly to the light vector mesons and reflect
their mass distribution. They are thus considered the prime
observable in studying mass (de-)generation related to restoration
of the spontaneously broken chiral symmetry. In order to draw
conclusions from the data (eg. CERES \cite{ce95}, NA60
\cite{Shahoyan:2005jj,Damjanovic:2005ni} or HADES) a precise theoretical description
of the medium effects has to be established.\\
During the early stages of a heavy ion reaction the system is out of equilibrium.
Also at later times, while the system further expands, it is (slightly) out of equilibrium.
In standard settings the system is thought to expand through quasi-equilibrium within a fireball scheme:
The dilepton production is then calculated under the assumption of instantaneous forms of the production rate at the given temperatures and convolution with the expansion \cite{ra96,Renk:2006ax}.  

In contrast, we account for possible retardation and off equilibrium by using a nonequilibrium quantum field
theoretical description based on the Kadanoff-Baym equations
\cite{kb62}. We simulate modifications of the light vector mesons
by introduction of a certain time dependence
of the self energies, which then allows for analyzing the dynamics
of the mesons' spectral properties as well as that of the
resulting dilepton rates. In \cite{Schenke:2005ry} we introduced time scales which
characterize the memory of the spectral function and compared them to typical
time scales in heavy ion reactions. We found that changes are not generally adiabatic and
memory effects have to be included. 
Here we present dilepton yields calculated for constant temperature 
to give a first impression of the memory effects.
Furthermore we concentrate on dilepton yields for a dropping mass scenario within a heavy ion collision
like that measured by the NA60 experiment \cite{Damjanovic:2005ni}.
Comparison of the experimental data with equilibrium calculations has lead many people to conclude that 
Brown-Rho scaling is ruled out by the data (see \cite{Brown:2005ka,Brown:2005kb} for a contrary opinion).  
We show that for this case memory effects are especially important and before drawing any conclusion they should be incorporated into any calculation.
 
\section{Equilibrium vs. Nonequilibrium}
\label{review}
 We compare the usual equilibrium approach, in that the dilepton
 rate is given by the well known formula
\begin{equation}
    \frac{dN}{d^4x d^4k}(\tau,k)=\frac{2 e^4}{(2\pi)^5}\frac{m_{\rho}^4}{g_{\rho}^2}\mathcal{L}(M)\frac{1}{M^2}n_{\text{B}}\left(T(\tau),k_0\right)\pi A_{\rho}
    (\tau,k)\text{,}
    \label{markovrate}
\end{equation}
 with $M^2=k^2=k_0^2-\mathbf{k}^2$ and $\mathcal{L}(M)=\left(1+\frac{2 m_l^2}{M^2}\right)\sqrt{1-\frac{4 m_l^2}{M^2}}\,\theta(M^2-4m_l^2)$,
 to the nonequilibrium formalism, in which the propagators of the
 $\rho$-meson and the virtual photon are calculated using the
 general nonequilibrium formulas and the dilepton rate for a time dependent but otherwise homogeneous situation is given
 by \cite{Schenke:2005ry}:
         \begin{align}
           \frac{dN}{d^4xd^4k}(\tau,k)=&2\frac{e^2}{(2\pi)^5}M^2\mathcal{L}(M)
           \Re\left[\int_{t_0}^{\tau}d\bar{t}\,i\,D_{\gamma\,T}^{<}(\textbf{k},\tau,\bar{t})e^{i
           k_0(\tau-\bar{t})}\right]\label{photrate1}\text{,}
        \end{align}
 with the transverse virtual photon propagator
 $D_{\gamma\,T}^{<}$, following the generalized fluctuation
 dissipation theorem
         \begin{equation}
            D^{<}=D^{\text{ret}} q^{<}D^{\text{adv}}\label{conv0}\text{,}
         \end{equation}
 with all integrals, time variables and further indices implicit. In this case the
 $D^{\text{ret/adv}}$ are the free retarded and advanced virtual photon
 propagators while $q^<=\Pi^<$, the photon self energy, which
 follows via VMD by
      $$
            \Pi^<_{T}(t_1,t_2)=e^2\frac{m_{\rho}^{*\,2}(t_1)}{g_\rho^*(t_1)} D_{\rho\,T}^{<}(t_1,t_2)\frac  {m_{\rho}^{*\,2}(t_2)}{g_\rho^*(t_2)}\text{.}\label{photvmd}
      $$
 Note that the couplings are to be taken at the correct vertices,
 separated in time.
 The meson propagator $D_{\rho\,T}^{<}$ also follows (\ref{conv0}),
 with $q^<=\Sigma^<$, the meson self energy, and the
 dressed retarded and advanced meson propagators which follow the
 equation of motion that in a spatially homogeneous and
 isotropic medium is given by
         \begin{align}
            \left(-\partial_{t_1}^2-m_{\rho}^{*\,2}-\textbf{k}^2\right)D_{\rho\,T}^{\text{ret}}(\textbf{k},t_1,t_2)-\int_{t_2}^{t_1}d\bar{t}
            \Sigma^{\text{ret}}_{\rho\,T}(\textbf{k},t_1,\bar{t})D_{\rho\,T}^{\text{ret}}(\textbf{k},\bar{t},t_2)=\delta(t_1-t_2)\text{,}\label{photdgl}
        \end{align}
 and
 $D_{\rho\,T}^{\text{adv}}(\textbf{k},t_1,t_2)=D_{\rho\,T}^{\text{ret}}(\textbf{k},t_2,t_1)$.
 The integral over past times in Eq.
 (\ref{photrate1}) as well as the time integrations in Eq.
 (\ref{conv0}) have encoded the finite memory of the system.
 In \cite{Schenke:2005ry} we show that for describing
 a heavy ion reaction it is essential to retain this
 dynamical information, because
 timescales for the adaption of the meson's spectral properties to
 the evolving medium are of the same order as the lifetime
 of the regarded hadronic system.
 Hence, an assumed adiabatic instantaneous 
 approximation as in Eq. (\ref{markovrate})
 can not properly describe the situation in a heavy ion collision,
 and remarkable differences in the resulting calculated yields
 occur.
 
 
\section{Nonequilibrium dilepton production from evolving media}
\subsection{Self energies} 
    The dynamic medium evolution is introduced by hand via a specified time dependent mass and retarded meson self
    energy $\Sigma^{\text{ret}}(\tau, \omega)$ \cite{Schenke:2005ry}. From that the self energy $\Sigma^<$
    follows by introducing a background temperature of the fireball.
    The fireball is assumed to generate the time dependent self energy $\Sigma^{\text{ret}}$ and, assuming a quasi thermalized system,
    the $\rho$-meson current-current correlator $\Sigma^<$ is given by
    \begin{equation}
     \Sigma^<_{\rho\,\text{T}}(\tau,\omega,\textbf{k})=2i n_{\text{B}}(T(\tau)) \text{Im} \Sigma^{\text{ret}}_{\rho\,\text{T}}(\tau,\omega,\textbf{k})\text{,}\notag
    \end{equation}
    which follows from the KMS relation, being valid for thermal systems. The latter is a rather strong assumption, but necessary in order to proceed.

The medium effects are introduced via a specific evolving retarded self energy
of the vector meson. A simple self energy
        \begin{align}
            \text{Im} \Sigma^{\text{ret}}(\omega,\tau)=-\omega\Gamma(\tau)\text{,}
        \end{align}
with a $\textbf{k}$- and $\omega$-independent width $\Gamma$,
leads to a Breit-Wigner distribution for the spectral function.
The time dependence is being accounted for by introduction of the
parameter $\tau$. For the $\textbf{k}=0$ mode, the full self
energy for coupling to $J^P=\frac{3}{2}^-$ -resonances is given by
        \begin{align}
            \text{Im}\Sigma^{\text{ret}}(\tau,\omega,\textbf{k}=0)=-\frac{\rho(\tau)}{3}
            \left(\frac{f_{RN\rho}}{m_{\rho}}\right)^2 g_I
            \frac{\omega^3 \bar{E}\Gamma_R(\tau)}{(\omega^2+\frac{\Gamma_R(\tau)^2}{4}-\bar{E}^2)^2+(\Gamma_R(\tau)\omega)^2}-\omega\Gamma(\tau)
            \label{modelshenself}\text{.}
        \end{align}
with $\bar{E}=\sqrt{m_R^2+\textbf{k}^2}-m_N$ and $m_R$ and $m_N$
the masses of the resonance and the nucleon respectively.
$\Gamma_R$ is the width of the resonance and $g_I$ the isospin
factor \cite{po04}. Other, more involved microscopic forms of the 
self energy $\text{Im}\Sigma^{\text{ret}}$ can also be invoked straight forwardly, see \cite{Schenke:2005ry}.

\subsection{Quantum interference}
\begin{figure}[htb]
  \hfill
  \begin{minipage}[htb]{.45\textwidth}
        \includegraphics[height=4.5cm]{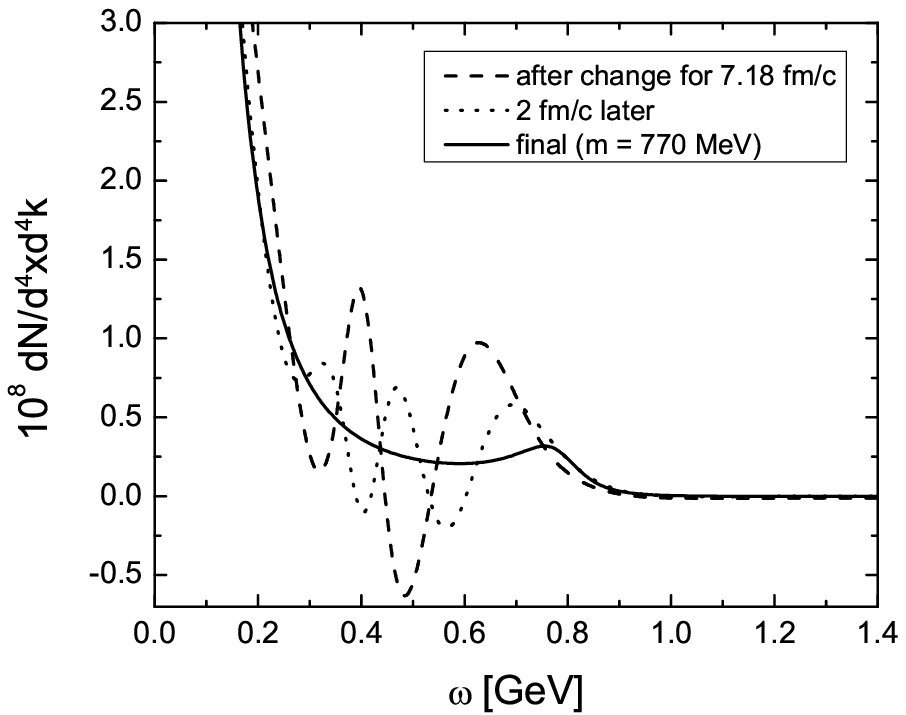}
        \caption{ Rate for the change of the mass from $m=400$ MeV to 770 MeV (constant $\Gamma$=150 MeV and constant $T=175$ MeV) directly after the self energy has reached its
                 final form (after 7.18 fm/c) and 2 fm/c later.}
        \label{fig:oscirate}
  \end{minipage}
  \hfill
  \begin{minipage}[htb]{.45\textwidth}
        \includegraphics[height=4.5cm]{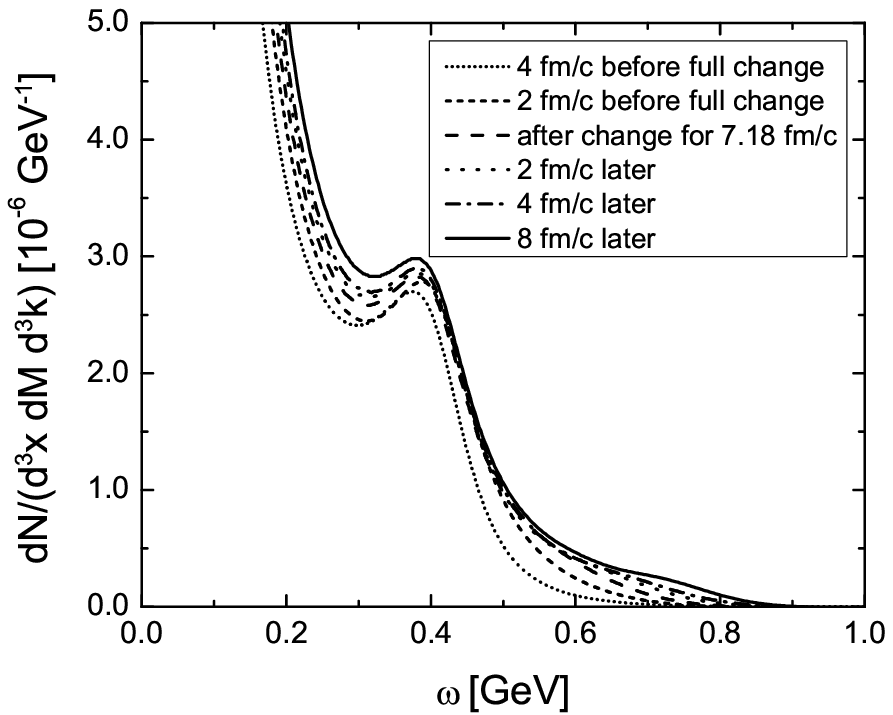}
        \caption{Yield corresponding to rates in the left plot integrated over time intervals starting with starting changes and ending after the indicated duration. Also yields at earlier and later times are shown.}
        \label{fig:osciyield}
  \end{minipage}
  \hfill
\end{figure}

If the systems properties change fast enough, as in relativistic heavy ion collisions, 
the quantum mechanical nature of the regarded systems can lead to
oscillations and negative values in the changing spectral
functions, occupation numbers and production rates as well as
interferences that one does not get in semi-classical, adiabatic
calculations. The rate calculated here possesses the full quantum
mechanical information incorporated and contains "memory"
interferences that might cause cancellations - hence the rate has
to be able to become negative while the time integrated yield
always stays positive as the only observable physical quantity. An
intriguing example for the occurring oscillations in electron-positron production is shown in
Figs. \ref{fig:oscirate} and \ref{fig:osciyield}.


\subsection{Yields at constant temperature}

In order to get a first impression of how the memory effects
affect dilepton yields, we investigate dielectron-yields from systems with a
constant size at constant temperature. 
Therefore we apply linear changes to the system over a duration of 7.18 fm/c.
The results for the yields integrated over this time for four
different scenarios are shown in Fig.
\ref{fig:constTV} below. The largest effect is found for the mass
shift of the $\rho$-meson: The yield is increased by a factor of
about 1.8 in the range from 200 to 450 MeV due to memory effects.
    \begin{figure}[htb]
      \begin{center}
        \includegraphics[height=9cm]{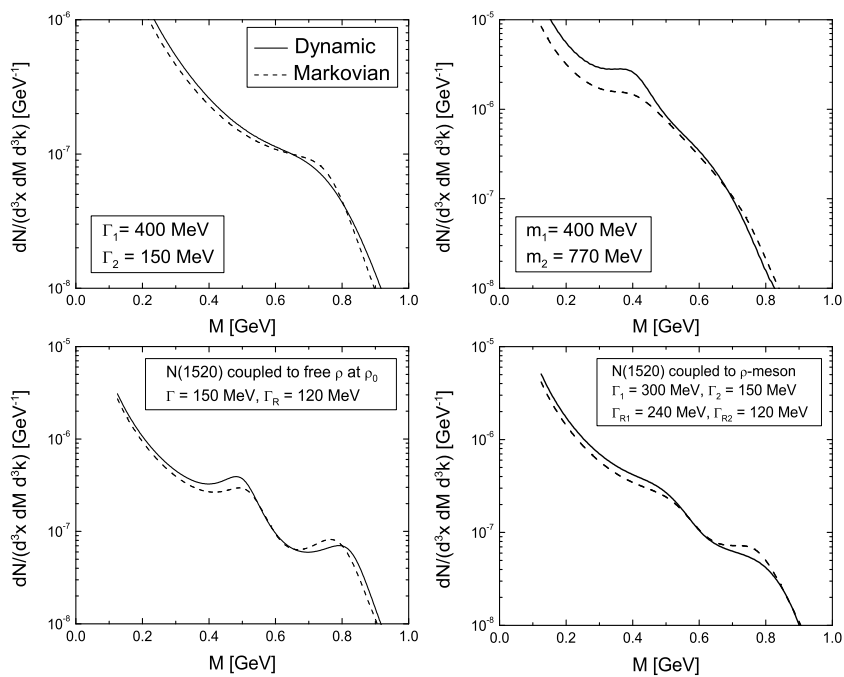}
        \caption{\small Comparison of the dynamically calculated (solid) to the Markovian (dashed) dilepton yields from an interval of
        duration $\Delta\tau=7.18$ fm/c in that the self energy was changed linearly as indicated in the corresponding figure.
        Initial quantities have index 1, final ones index 2. The temperature was kept constant at $T=175$ MeV.}
        \label{fig:constTV}
      \end{center}
    \end{figure}
The other cases show differences, but not as pronounced - we see a
minor enhancement of the lower mass tail due to the Bose factor in
all cases. In the case of a broadened $\rho$-meson, one can
nicely see that the dynamically calculated yield is a broader
distribution and that for the case of the coupling to the N(1520)
without broadening the resonance peak is stronger, due to the
system's memory of it, and the vacuum peak is shifted further to
the right due to memory of the level repulsion effect.

To conclude this section we state that for the $\rho$-meson we
found exactly the modifications of the yields that one would
expect qualitatively when including memory effects for the
constant temperature case. In a dynamical fireball setup with full decoupling in the end
we find similar results but more pronounced differences between the quasi-equilibrium and
nonequilibrium calculations \cite{Schenke:2005ry}.


\section{Dilepton yields from heavy ion reactions}
\label{fireball}
 Triggered by the recent discussion on whether Brown-Rho scaling
 \cite{br91}
 is 'ruled out' by the NA60 data, presented at Quark Matter 2005 \cite{Damjanovic:2005ni},
 we concentrate on this issue and calculate dilepton yields for both suggested mass
 modifications of the $\rho$-meson within the nonequilibrium
 formulation, which has been introduced in \cite{Schenke:2005ry}.
 First quantitative but simplified calculations within this work
 have shown that memory effects, which are neglected in
 equilibrium calculations (recent calculations for the NA60 scenario can be found in \cite{vanHees:2006ng} and \cite{Renk:2006ax}), have an important influence
 on the final dilepton yields, especially for dropping mass scenarios.
 An analysis of which scenario describes the data correctly demands
 the proper inclusion of memory effects. The validity of
 equilibrium calculations will depend on the strength of these
 effects.
 The two mass parameterizations we use are the
 temperature and density dependent one used by Rapp et al. \cite{rapp1}
 \begin{equation}
    m_{\rho}^{*}=m_{\rho}(1-0.15
    n_{\text{B}}/n_0)\left[1-(T/T_{\text{c}})^2\right]^{0.3}\text{,} \label{rapppar}
 \end{equation}
 assuming a constant gauge coupling $g$ in the vector meson dominance (VMD) -coupling,
 and one motivated by an improved version of Brown-Rho scaling \cite{Brown:2005ka,Brown:2005kb}
 , where
 \begin{equation}
    m_{\rho}^{*}=m_{\rho}(1-0.15 n_{\text{B}}/n_0)\text{,}\label{brownpar}
 \end{equation}
 with a modified gauge coupling $g^*$, in such a way that $g^*$ is
 constant up to normal nuclear density $n_0$, while from then on
 $m^*_{\rho}/g^*$ is taken to be constant \cite{Brown:2002is}.
 This is due to the fact that VMD, which accounts for most of the
 shape of the dilepton spectrum, is violated for most
 temperature and density regions (see \cite{Brown:2005ka,Brown:2005kb} and
 \cite{Harada:2003jx}). Brown and Rho argue \cite{Brown:2005ka,Brown:2005kb} that
 because lattice calculations tell us that
 the pole mass of the vector mesons does not change
 appreciably up to $T=125$ MeV, the parameterization using
 $[1-(T/T_{\text{c}})^n]^{d}$ (with positive $d$ and integer $n$)
 overestimates the mass shift. They find temperature dependent effects to be
 an order of magnitude smaller than the density dependent
 effects, and hence suggest to concentrate on the density
 dependent part. Furthermore they point out that due to the
 violation of VMD the overall dilepton production in dense matter
 should be reduced by a factor of 4 compared to Rapp's
 calculations, which we do not take into account here.
    Following \cite{Rapp:1999us,vanHees:2006ng}, we use a fireball model with a cylindrical volume expansion
    in the $\pm z$ direction:
    \begin{equation}
        V(\tau)=(z_0+v_z \tau)\pi(r_0+0.5\,a_{\perp}\tau^2)^2\text{,}
    \end{equation}
    where $z_0$ is equivalent to a formation time $\tau_0=1$ fm/c, $v_z=c$ is the
    primordial longitudinal motion, $r_0=5.15\,\text{fm}$ is the initial nuclear overlap radius,
    and $a_{\perp}=0.08\,c^2/\text{fm}$ is the radial acceleration.
    We start the calculation at $T=T_c=175$ MeV and set the meson self energy
    $\Sigma^<$ to zero for temperatures below chemical freezeout at $T=120$ MeV.
    The hadronic phase lives for about $6$ fm/c.
 The calculation is done for each momentum mode between $k=0$ and $k=1.5$ GeV and integrated over momenta.
 
\begin{figure}[htb]
  \hfill
  \begin{minipage}[htb]{.45\textwidth}
      \begin{center}
        \includegraphics[height=4cm]{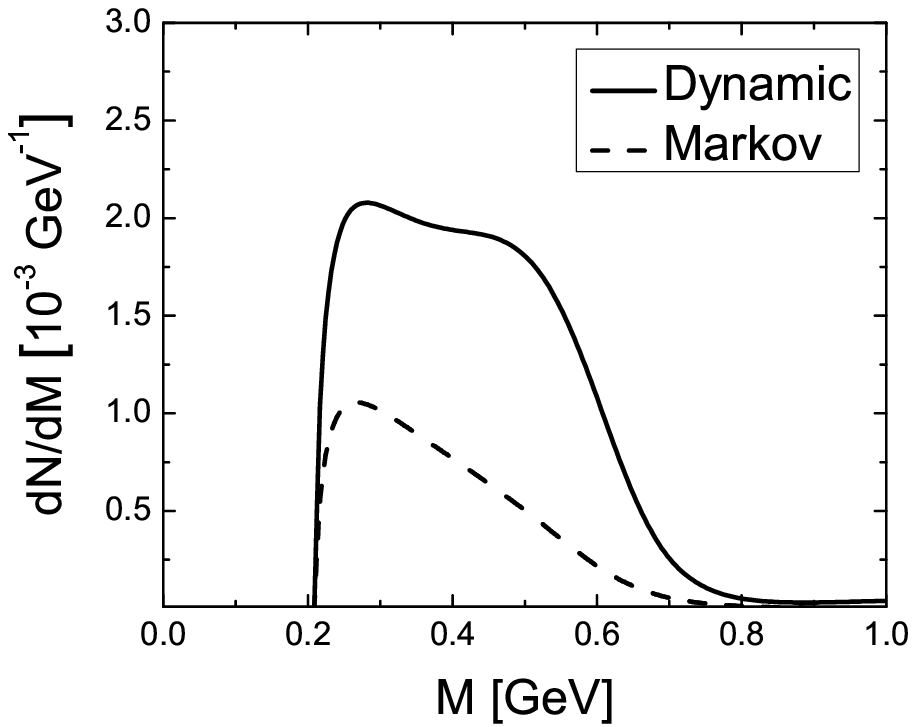}
        \caption{Dimuon yield from decaying $\rho$-mesons calculated out of equilibrium and within the usual instantaneous approximation using parameterization (\ref{rapppar}).
        A difference of about a factor of 4 is clearly visible in the regime around 500 MeV.}
        \label{fig:rapp}
      \end{center}
  \end{minipage}
  \hfill
  \begin{minipage}[htb]{.45\textwidth}
      \begin{center}
        \includegraphics[height=4cm]{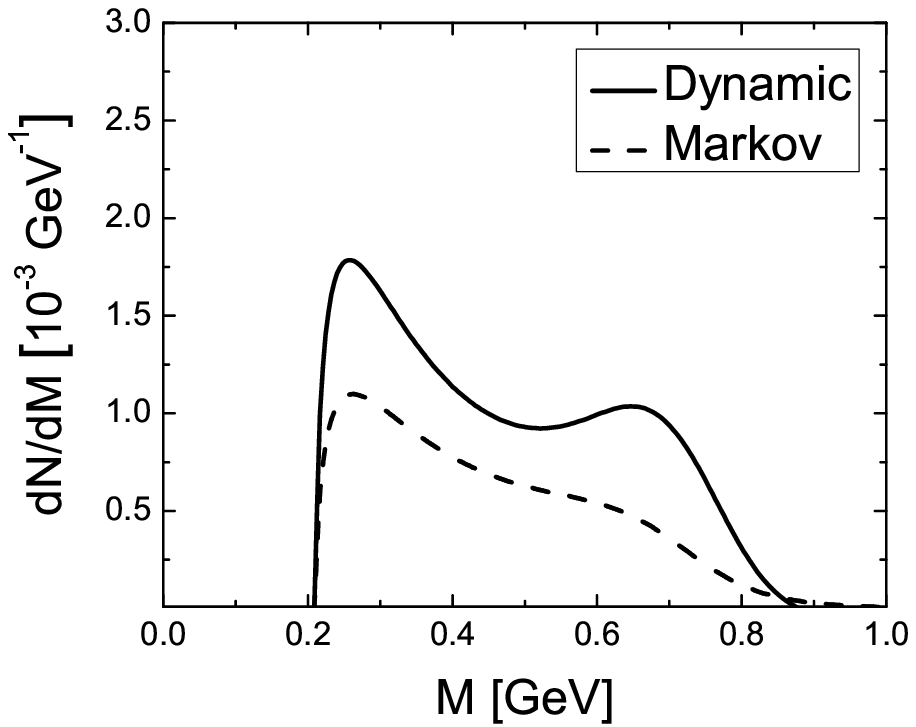}
        \caption{Dimuon yield from decaying $\rho$-mesons calculated out of equilibrium and within the usual instantaneous approximation using parameterization (\ref{brownpar}).
        The most significant difference (about a factor of 3) is visible between 600 and 700 MeV.}
        \label{fig:br}
      \end{center}
  \end{minipage}
  \hfill
\end{figure}

 The results for the dimuon spectra are shown in Figs. \ref{fig:rapp} and \ref{fig:br}. We find a strong enhancement and also modification of the shape in both cases where for the density and temperature dependent parameterization the dynamic calculation yields about a factor of 4 more dileptons around an invariant mass of 500 MeV as compared to the static case, while the other parameterization shows more enhancement ($\sim \times 3$) around 650 MeV. \\
The reason for the strong differences are the following: First, effectively the meson mass approaches its vacuum value more slowly in the dynamic calculation. Lower masses are enhanced due to the Bose factor and if the spectral function has a lower mass for a longer time this means more enhancement. Furthermore does the memory of higher temperatures increase this enhancement for all masses. Finally, the modified coupling in Eq. (\ref{photvmd}) suppresses the early contributions. That means that for the static case all low masses in the spectral function and high temperatures are suppressed whereas in the dynamic case these low masses and high temperatures still contribute when the couling is larger and hence they are less suppressed by the VMD-coupling.

\section{Summary and Conclusions}
\label{conclusion}
    We calculated dilepton
    yields within a non-equilibrium field theory formalism,
    based on the Kadanoff-Baym equations.
    We investigated possible medium modifications of the $\rho$-meson
    for constant temperature and 'Brown-Rho'-scaling within an evolving fireball.
    Special attention was put to possible retardation effects
    concerning the off-shell evolution of the vector mesonic excitations.

    The full quantum field theoretical treatment leads to
    oscillations in all mentioned quantities when changes in the self energy are performed.
    This oscillatory behavior reveals the quantum mechanical
    character of the many particle system, present in the investigated heavy ion reaction.

    Comparison of dynamically calculated yields with those calculated assuming adiabaticity reveals moderate to strong differences.
    For the dropping mass scenarios investigated, an enhancement of about a factor of 3-4 below the $\rho$-vacuum mass and significant differences in shape have been found, leading to the conclusion that memory effects must not be neglected in precision calculations of dilepton yields from relativistic heavy ion collisions.

%


\bibliography{lajolla2006-greiner}
\bibliographystyle{lajolla2006}
 
\vfill\eject
\end{document}